\documentclass[aps,twocolumn,showpacs,superscriptaddress,preprintnumbers,
amsmath,amssymb,floatfix,pre]{revtex4}  
\usepackage{graphicx}  
\usepackage{dcolumn}   
\usepackage{bm}        
\usepackage{amsmath}   
\usepackage{amsfonts}
\usepackage{amssymb}
\usepackage{natbib}
\usepackage{color}

\begin{document}

\title{Plastic response and correlations in athermally sheared amorphous solids}
\author{F. Puosi}
\email{francesco.puosi@univ-grenoble-alpes.fr}
\affiliation{Univ. Grenoble Alpes, LIPHY, F-38000 Grenoble, France}
\affiliation{CNRS, LIPHY, F-38000 Grenoble, France}

\author{J. Rottler}
\email{jrottler@phas.ubc.ca}
\affiliation{Department of Physics and Astronomy, The University of British Columbia,6224 Agricultural Road, Vancouver, British Columbia V6T 1Z1, Canada}

\author{J-L. Barrat}
\email{jean-louis.barrat@univ-grenoble-alpes.fr}
\affiliation{Univ. Grenoble Alpes, LIPHY, F-38000 Grenoble, France}
\affiliation{CNRS, LIPHY, F-38000 Grenoble, France}
\affiliation{Institut Laue-Langevin, 6 rue Jules Horowitz, BP 156, F-38042 Grenoble, France}
\date{\today}

\begin{abstract}
The onset of irreversible deformation in low-temperature amorphous solids is due to the accumulation of elementary events, consisting of spacially and temporally localized atomic rearrangements involving only a few tens of atoms. Recently, numerical and experimental work addressed the issue of spatio-temporal correlations between these plastic events.  Here, we provide further insight into these correlations by investigating, via molecular dynamics (MD) simulations,  the plastic response of a two-dimensional amorphous solid to artificially triggered local shear transformations. We show that while the plastic response is virtually absent in as-quenched configurations, it becomes apparent if a shear strain was previously imposed on the system.  Plastic response has a four-fold symmetry which is characteristic of the shear stress redistribution following the local transformation. At high shear rate we report evidence for a fluctuation-dissipation relation, connecting plastic response and correlation, which seems to break down if lower shear rates are considered. 


\end{abstract}

\pacs{}
\maketitle

\section{\label{sec:intro}Introduction}

Heterogeneity is a crucial aspect of the flow of amorphous materials. If these material are driven by an external shear, one observes localized particle rearrangements, called shear transformations (STs), taking place in a small region while the rest of the system deforms elastically \cite{Argon1979, FalkPRE57, Schall07}. The effect of a shear transformation, i.e. the stress redistribution in the surrounding medium,  is usually described via an elastic propagator $\mathcal{G}$, which is the solution of the Eshelby inclusion problem in an uniform elastic medium \cite{Eshelby57}.  In two dimensions, $\mathcal{G}$ has a quadrupolar symmetry and it decays as $r^{-2}$ in space. The elastic propagator is the key ingredient of rheological models for the flow of amorphous materials \cite{PicardEPJE04, PicardPRE05, VandemPRB11, MartensPRL11, MartensSM12, NicolasSM14}. 

In a recent paper \cite{PuosiSTdyn}, we addressed via computer simulations of a model amorphous solid the question of the \textit{elastic} response of an amorphous solid to localized shear transformations. We showed that the Eshelby description holds on average while for individual plastic events the response is blurred by strong fluctuations, presumably associated with the elastic heterogeneity of the material. In order to capture these  fluctuations within coarse-grained rheological models, it is necessary to go beyond the equilibrium-based description of the elastic propagator  \cite{Nicolas2015333}. 

Here, we extend our previous results and investigate \textit{plastic} effects due to STs in amorphous solids. The question we want to address is to what extent a ST is able to induce subsequent plastic events in the surrounding regions. This goes back to the topic of plastic correlations, namely how a plastic event is influenced by the position of the events that occurred in the past. Plastic correlations have the strong potential to provide  information about the dynamical organization of the plastic flow. This possibility motivated recent work on this topic. In athermal quasi-static simulations, Maloney and Lema\^{\i}tre showed that elementary events tend to organize into correlated avalanches  \cite{MaloneyPRL93_b}.  Later, evidence for a correlation in the non-affine displacements of a colloidal glass was first reported by Chikkadi and co-workers in experiments \cite{ChikkadiEPL12} and then confirmed by Mandal and Varnik \cite{MandalPRE13,VarnikPRE14} in numerical simulations. In Ref. \onlinecite{LemaitrePRL13} correlations of the local strain field were found to emerge at the transition between the Newtonian and shear-thinning regime in a flowing liquids.  Similarly ,in  Ref. \onlinecite{BenziSM14}, the authors reported correlated plastic events in a simulation of a concentrated emulsion. Recently, some of us showed how a simple coarse-grained model is able to reproduce, with some small quantitative discrepancies, the spatio-temporal correlations between plastic events in the flow of a disordered athermal solid \cite{NicolasEPJE14}.  

In this work, we investigate plastic correlations in a model amorphous solid, following the approach of Ref.~\onlinecite{PuosiSTdyn} which consists of inducing artificial shear transformations in the system and observing their response in time. We will propose a detailed description of the plastic response and compare it with previous correlation results. The paper is organized as follows. Details about the model and the procedure to simulate artificial shear transformations are given in Sec. \ref{sec:methods}. Results of the numerical simulations are discussed in Sec. \ref{sec:results}, while the final Sec. \ref{sec:con} provides a short summary and discussion.

\section{Model and details of the simulation}\label{sec:methods}

We consider a generic two-dimensional (2D) model of a glass, consisting of a mixture of A and B particles, with $N_A=32,500$ and $N_B=17,500$, interacting via a Lennard-Jones potential 
  $ V_{\alpha\beta}(r)= 4 \epsilon_{\alpha \beta}\left[ \left( \frac{\sigma_{\alpha\beta}}{r} \right)^{12} - \left( \frac{\sigma_{\alpha\beta}}{r} \right)^{6}  \right ] $
with $\alpha,\beta=A,B$ and $r$ being the distance between two particles.  The parameters $\epsilon_{AA}$, $\sigma_{AA}$ and $m_A$ define the units of energy, length and mass; the unit of time is given by $\tau_0=\sigma_{AA}\sqrt{(m_A/\epsilon_{AA})}$. We set $\epsilon_{AA}=1.0 $, $\epsilon_{AB}=1.5 $, $\epsilon_{BB}=0.5 $, $\sigma_{AA}=1.0$, $\sigma_{AB}=0.8$ and $\sigma_{BB}=0.88$ and $m_A=m_B=1$. With this choice, the system is stable against crystallization in two dimensions \cite{Kob2d09}.  A similar system was used by Falk and Langer \cite{FalkPRE57} to study plasticity in 2D metallic glasses. The potential is truncated at $r=r_c=2.5$ for computational convenience. The simulation box dimensions $L_x=L_y=205$ are fixed and periodic boundary conditions are used. The equations of motion are integrated using the velocity Verlet algorithm with a time step $\delta t=0.005$. 
The athermal limit is achieved by thermostating the system at zero temperature via a Langevin thermostat \cite{LangevinThermo} with a damping coefficient $\Gamma=1$; the associated equations of motion are:
\begin{eqnarray}
\frac{d\mathbf{r}_i}{dt}&=&\frac{\mathbf{p}_i}{m} \\
\frac{d\mathbf{p}_i}{dt}&=&-\sum_{j\neq i}\frac{\partial V(\mathbf{r}_{ij})}{\partial \mathbf{r}_{ij}}-\Gamma\mathbf{p}_i  
\end{eqnarray}
where $(\mathbf{p}_i,\mathbf{r}_i)$ are the momentum and the position of particle $i$. As $T=0$, no fluctuating force appears in the equations. 

Glassy states were prepared by quenching  equilibrated systems at $T=1$ to zero temperature with a fast rate $dT/dt=2\times 10^{-3}$ while maintaining constant volume. Simple shear is imposed at a rate $\dot{\gamma}$ by deforming the box dimensions and remapping the particles positions. Local shear transformations (STs) are generated by applying a pure shear strain $\epsilon$ to a circular region of radius $a=2.5$ centered at $(x_0,y_0)$,  as discussed in Ref. \onlinecite{PuosiSTdyn}.  Particles inside the region, at the initial position $(x_i,y_i)$ are displaced to $(x_i',y_i')$  according to:
\begin{equation} \label{eqn:ST}
\left\{ 
 \begin{array}{lr}
  x_i\rightarrow x'_i=x_i+\epsilon (y_i-y_0) \\
  y_i\rightarrow y'_i=y_i+\epsilon (x_i-x_0) 
 \end{array}
\right.
\end{equation}
The transformation is instantaneous and sets the time origin $t=0$. The positions of the particles in the ST are frozen and the behavior of the surrounding ones at later times is observed. For as-quenched configurations, the response is averaged over $10$ independent realizations and for each of those $50$ position for the ST center $(x_0,y_0)$ are considered. For pre-sheared configurations, we average the response over independent realizations ($4$ starting configurations), strain ($16$ strain values in the range $0.2\leq\gamma\leq 1.0$) and position of the ST center (20 position), resulting in an average over $1280$ trajectories. 

Plastic activity is described by the $D^2_{min}$ quantity introduced by Falk and Langer \cite{FalkPRE57}, which evaluates deviations from an affine deformation on a local scale. For a given particle $i$,  $D^2_{min}$ is defined as the minimum over all possible linear deformation tensors $\mathbf{\epsilon}_{loc}$ of:
\begin{equation}\label{eqn:d2min}
D^2(i,t_0, t)=\sum_j \left [ r_{ij}(t_0+ t) - \left ( \mathbb{I} + \mathbf{\epsilon}_{loc} \right) r_{ij}(t_0) \right]^2
\end{equation}
where the index $j$ runs over all the neighbors of the reference particle $i$ and $\mathbb{I}$ is the identity matrix. 

\section{\label{sec:results}Results and discussion}

\subsection{As-quenched configurations}

\begin{figure}[t]
\begin{center}
\includegraphics[width=1.\columnwidth]{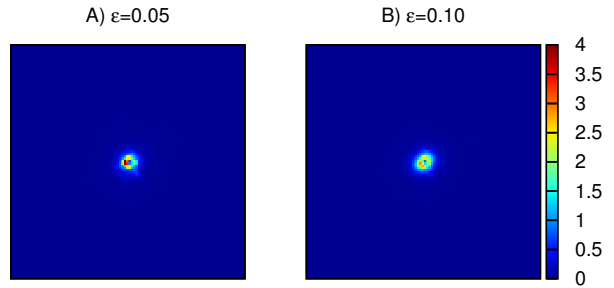}
\end{center}
\caption{Plastic response field $R_2(\mathbf{r},\Delta t)$ induced by a shear transformation (at the center of the cell) in the long time limit $\Delta t=10^3$ for as-quenched configurations. A region of size $200\times200$ around the ST is shown and the color corresponds to the amplitude of the plastic response field.}
\label{fig1}
\end{figure}

In Ref. \onlinecite{PuosiSTdyn}, the local strain $\epsilon$ was set to a few percent ($\epsilon=0.025$) in order to probe the elastic reversible response of the system to STs. Here, we investigate the effects of a higher local strain in amorphous configurations at $T=0$, rapidly quenched from high temperature. We define the plastic response function $R_2(\mathbf{r},\Delta t)=\overline{{D}^2}_{min}(\mathbf{r},t_0, t_0+\Delta t)/(a^2 \epsilon^2)$ where $\overline{{D}^2}_{min}$ is the coarse-grained $D^2_{min}$-field obtained by mapping particles into a grid with a spacing $\xi=1$, $t_0$ is the time at which the ST is applied and $a$ and $\epsilon$ are the radius and local strain of the ST, respectively. In Fig.~\ref{fig1} spatial maps of $R_2(\mathbf{r},\Delta t)$ for $\Delta t=10^3$, which corresponds to the long time limit, are shown for larger values of the applied strain. We observe that the plastic activity is very weak, being restricted to a small region close to the ST center, and that it does not depend on the specific value of the strain $\epsilon$. Even for a strain of 10\%, the response is almost entirely elastic. 

\subsection{Sheared configurations}

\begin{figure}[h!]
\begin{center}
\includegraphics[width=1\columnwidth]{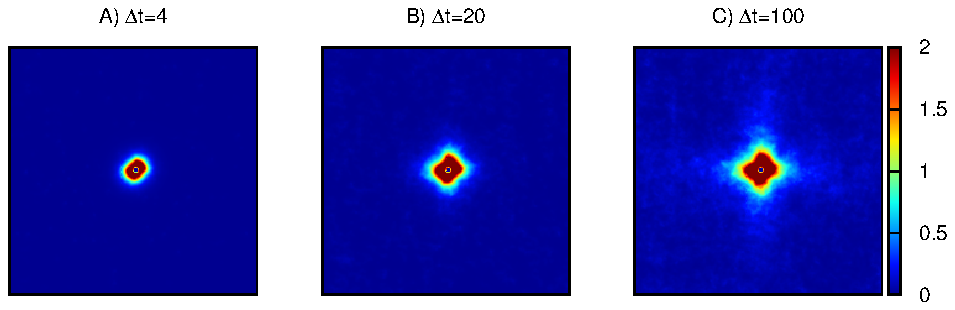}\vspace{-0.5cm}
\includegraphics[width=1\columnwidth]{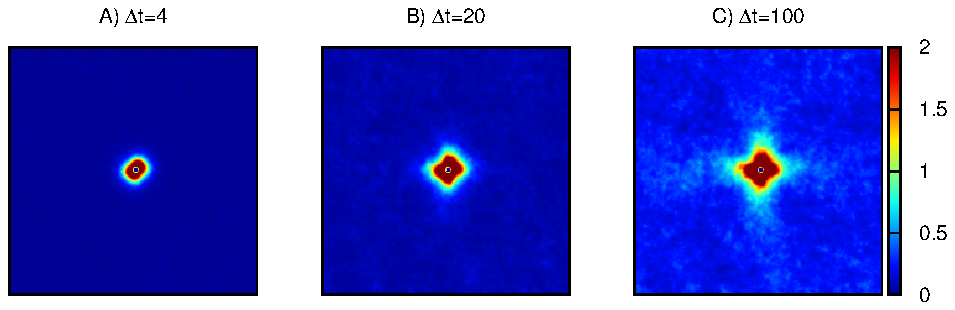}\vspace{-0.5cm}
\includegraphics[width=1\columnwidth]{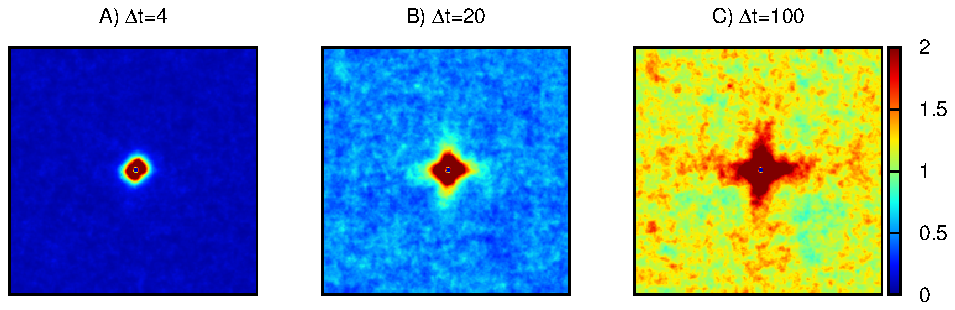}
\end{center}
\caption{Plastic response field $R_2(\mathbf{r},\Delta t)$ induced by a ST in sheared configurations for different time lags $\Delta t$. Data are shown for different shear rates $\dot{\gamma}=10^{-6}$ (top), $\dot{\gamma}=10^{-5}$ (middle), and $\dot{\gamma}=10^{-4}$ (bottom).  A region of size $200\times200$  around the ST is shown. }
\label{fig2} 
\end{figure}

We now focus on the effects of external deformation on the tendency of the system to undergo plastic rearrangements. Simple shear is imposed on the system at a rate $\dot{\gamma}$ before applying the shear transformation protocol. Steady state configurations with a strain $\gamma\geq 0.2$ are considered as starting configurations. Then the accumulated stress is not relaxed and the strain $\gamma$ is kept constant in the following temporal evolution, specifically $\dot{\gamma}$ is instantaneously set to zero. The time evolution of the plastic response in sheared configurations is shown in Fig.~\ref{fig2} for three distinct shear rates $\dot{\gamma}=10^{-6}$, $10^{-5}$ and $10^{-4}$.  
First, with respect to the case of as-quenched systems, the effect of the pre-shearing is apparent in the increased plastic activity even for short time lags. Looking at long times, the pattern of the plastic activity clearly resembles the elastic propagator $\mathcal{G}\sim \cos{(4\theta)}/r^2$, which controls the stress redistribution following the ST. High plastic intensity is observed in the streamwise ($\theta=0^{\circ}$) and crosswise ($\theta=90^{\circ}$) directions, which correspond to the directions of positive stress release. On the other hand, the redistributed stress is negative along the diagonal and this results in lower plastic activity. It is straightforward to rationalize these observations: in the pre-sheared configurations, many regions have already been loaded close to the yield point, and the stress redistribution following the primary ST can trigger  plastic events much more easily than starting from as-quenched configurations. 

\begin{figure}[t]
\begin{center}
\includegraphics[width=0.95\columnwidth]{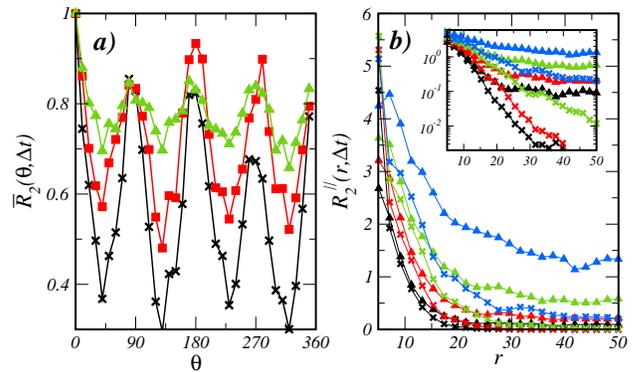}
\end{center}
\caption{a): angular dependence of the plastic response $\overline{R}_{2}(\theta, \Delta t)=\alpha \int_{2a}^{L/2} R_{2}(r, \theta, \Delta t) dr$ (see text). Data are shown for $\Delta t=100$ and for different shear rates $\dot{\gamma}=10^{-4}$ (triangles), $\dot{\gamma}=10^{-5}$ (squares) and $\dot{\gamma}=10^{-6}$ (crosses). b): plastic response along the shear direction $R_2^{\|}$ in lin-lin (main panel) and lin-log (inset) plots. Data are shown for different time lags $\Delta t=4$ (black), $\Delta t=8$ (red), $\Delta t=20$ (green) and $\Delta t=40$ (blue), and for  different shear rates $\dot{\gamma}=10^{-4}$ (triangles) and $\dot{\gamma}=10^{-6}$ (crosses). }
\label{fig3}
\end{figure}

This picture becomes clearer if one considers the angular dependence of the plastic response, as denoted by the quantity $\overline{R}_{2}(\theta, \Delta t)=\alpha \int_{2a}^{L/2} R_{2}(r, \theta, \Delta t) dr$  where $L$ is the system size and the prefactor $\alpha$ is chosen such that $\overline{R}_{2}(\theta, \Delta t)$ has a maximum of $1$. The long time limit of $\overline{R}_{2}(\theta, \Delta t)$ is shown in Fig.~\ref{fig3}(a) for different shear rates. 
We observe again the quadrupolar modulation characteristic of the Eshelby response function, which becomes more pronounced with decreasing shear rate. Here we point out that no clear asymmetry is observed between streamwise and crosswise lobes, in contrast with previous results on plastic correlations \cite{NicolasEPJE14}, where streamwise peaks appeared to be stronger than crosswise ones at low shear rates. However, we cannot exclude that in our case this effect is still hidden by the noise. Fig. \ref{fig2} and Fig. \ref{fig3} also show that as $\dot{\gamma}$ increases a plastic background emerges. This is due to a cascade effect, with the primary ST triggering plastic rearrangements which themselves play the role of sources for subsequent events.  This phenomenon is more prevalent at higher driving rates. 

In Fig. \ref{fig3}b) we show the spatial decay of the response function along the shear direction $R_2^{\|}$ at different times and for different shear rates. The response extends to larger distances as the time interval increases, consistent with the propagation of the strain field created by the ST, as discussed previously \cite{PuosiSTdyn}.  The decay of $R_2^{\|}$ is approximately exponential (see the inset of the figure).  This agrees with the observation of an exponential decay in the correlations of plasticity for models of amorphous systems implementing a ``mean-field" dissipation scheme in the equation of motions \cite{VarnikPRE14, NicolasEPJE14}, similar to the one used in the present work. Further, the response seems to be independent on the shear rate: at short times data for the different shear rates are indistinguishable whereas deviations are observed at larger times due to the emerging background plasticity mentioned before. 

\begin{figure}[th]
\begin{center}
\includegraphics[width=1\columnwidth]{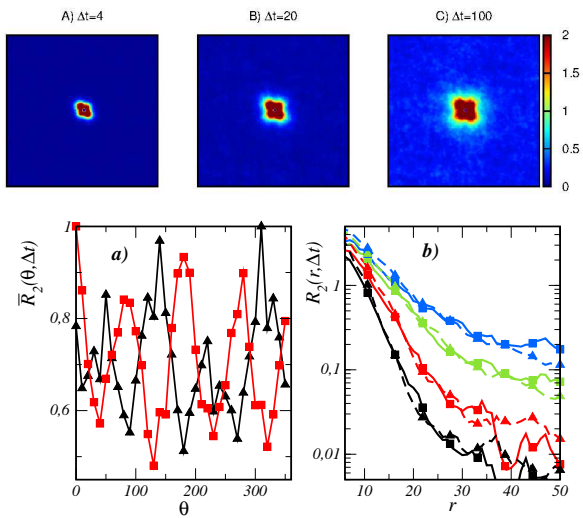}
\includegraphics[width=0.85\columnwidth]{figure423n.eps}

\end{center}
\caption{Top: plastic response field $R_2(\mathbf{r},\Delta t)$ at different time lags induced by a ST whose main axis is rotated by an angle $\phi=90^{\circ}$ with respect to the shear direction ($\dot{\gamma}=10^{-5})$. Bottom, panel a): corresponding angular dependence for $\Delta t=100$ (triangles); for comparison, data for $\phi=0^{\circ}$ are also shown (squares). Bottom, panel b): spatial decay along a principal direction $\theta^*$ at different time lags, $\Delta t=4$ (black), $\Delta t=8$ (red), $\Delta t=20$ (green) and $\Delta t=40$ (blue), for $\phi=90^{\circ}$ with $\theta^*=45^{\circ}$ (triangles) and the reference $\phi=0^{\circ}$ with $\theta^*=0^{\circ}$ (squares). }
\label{fig4}
\end{figure}

In order to confirm the proposed scenario of a plastic response controlled by the Eshelby elastic propagator, we investigate the response to a ST whose principal axes are rotated by an angle $\phi=90^{\circ}$ with respect of the shear direction. This rotation of the local strain matrix is equivalent to a sign change in  $\epsilon$. According to the quadrupolar symmetry of the propagator, this should result in a rotation of the response pattern by $45^{\circ}$ with respect to the one observed for $\phi=0^{\circ}$ discussed above. In Fig. \ref{fig4} we show the response to the rotated ST. If we focus on the angular dependence (panel a), we observe that the main peaks are shifted by $45^{\circ}$, as expected. Further, as shown in the panel b) of Fig. \ref{fig4}, the spatial decay of the response at different time lags along the principal direction $\theta^*=45^{\circ}$ is in very good agreement with that for the unrotated case where the shear direction is $\theta^*=0^{\circ}$.

\subsection{Response and correlation} 

We now turn to a quantitative comparison between plastic response and correlations. In Ref.~\onlinecite{NicolasEPJE14}, some of us presented for the same model amorphous solid a detailed description of the plastic events and their dynamical correlations, resolved both in space and time, using the two-point, two-time plastic correlator
\begin{multline}
C_2(r,\Delta t)= \alpha \left ( \left \langle \overline{D^2_{min}(r_0,t_0)D^2_{min}(r_0+r,t_0+\Delta t)} \right \rangle \right.  \\- \left . \left \langle \overline{D^2_{min}(r_0,t_0)D^2_{min}(r_0,t_0+\Delta t)} \right \rangle  \right ),
\end{multline}
where the brackets denote an average over time $t_0$, the bars represent an average over spatial coordinate $r_0$ and the prefactor $\alpha$ is chosen such that $C_2(r=0,\Delta t=0)=1$.  $C_2$ measures   the (enhanced or reduced) likelihood that a plastic event occurs at $r_0+r$ if a plastic event was active at position $r_0$ some prescribed time $\Delta t$ ago. 

In the first instance, one could imagine to compare directly the correlation $C_2$ to the response function $R_2$. In this case, the two quantities show strongly different behavior with the correlation extending to larger distances with respect to response for equal time lags (not shown). However, we argue that this in not the most significant comparison.  Indeed, assuming a linear response perspective, with the strain $\epsilon$ of the ST acting as perturbation, the most appropriate quantity to focus on seems to be $(D^2_{min})^{1/2}$ rather than $D^2_{min}$ ($D^2_{min}$ is a squared displacement, which depends quadratically on the local strain tensor $\mathbf{\epsilon}_{local}$). In this spirit, we define the response function $R_1(r,\Delta t)$:
\begin{equation} \label{eqn:r1}
R_1(r,\Delta t)=\left(R_2(r,\Delta t)-R_2^{\infty}(\Delta t)\right)^{1/2}
\end{equation}
and the corresponding correlation function $C_1(r,\Delta t)$:
\begin{multline} \label{eqn:c1}
C_1(r,\Delta t)=  \alpha \left ( \left \langle \overline{(D^2_{min}(r_0,t_0))^{1/2}(D^2_{min}(r_0+r,t_0+\Delta t))^{1/2}} \right \rangle \right.  \\- \left . \left \langle \overline{(D^2_{min}(r_0,t_0))^{1/2}(D^2_{min}(r_0,t_0+\Delta t))^{1/2}} \right \rangle  \right )
\end{multline}
where in Eq. \ref{eqn:r1} the background response $R_2^{\infty}(\Delta t)$ is subtracted. 

\begin{figure}[t]
\begin{center}
\includegraphics[width=0.95\columnwidth]{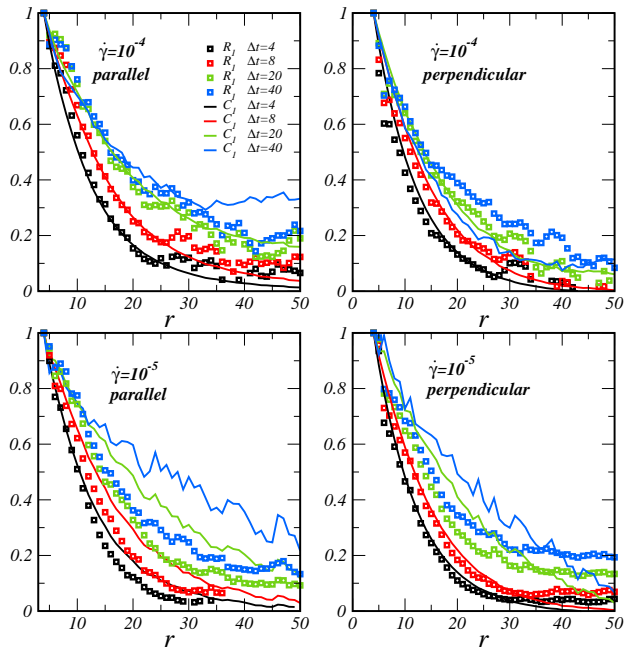}
\end{center}
\caption{Decay of the plastic response $R_1(r,\Delta t)$ (symbols) and correlation  $C_1(r,\Delta t)$ (lines) in the direction parallel and perpendicular to the shear direction. Data are shown for different shear rates $\dot{\gamma}=10^{-4}$ (top panels), $\dot{\gamma}=10^{-5}$ (bottom panels) and for different time lags $\Delta t=4$ (black), $\Delta t=8$ (red), $\Delta t=20$ (green) and $\Delta t=40$ (blue). Data are vertically rescaled in order to have $R_1(r,\Delta t)=C_1(r,\Delta t)=1$ at $r=4$.  }
\label{fig5}
\end{figure}

In Figure \ref{fig5}, we compare the decay of the response and correlation functions $R_1(r,\Delta t)$ and $C_1(r,\Delta t)$ along the directions parallel and perpendicular to the shear direction. First, we focus on the high shear rate $\dot{\gamma}=10^{-4}$. Here, we observe that the decay of the response strongly resembles that of the correlation showing an almost exponential decay with a comparable extension. This applies both to the parallel and perpendicular directions. By contrast, for the low shear rate $\dot{\gamma}=10^{-5}$ differences become apparent, with the correlation extending to larger distances. Deviations are stronger in the shear direction, especially for large time lags, whereas they are weaker in the perpendicular direction. 

\begin{figure}[th]
\begin{center}
\includegraphics[width=0.9\columnwidth]{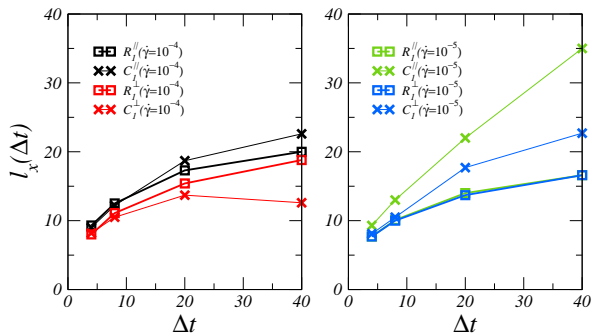}
\end{center}
\caption{Time dependence of the decay length $l$ for the plastic response (open squares) and correlation (crosses). Data are shown for the directions parallel and perpendicular to the shear direction and for different shear rates. }
\label{fig6}
\end{figure}

Assuming purely exponential behavior $x(r,\Delta t)= A_{x}(r,\Delta t)\exp(-r/l_x(\Delta t))$  with $x=R_1,C_1$, we can estimate the decay length $l_x(\Delta t)$. In Fig. \ref{fig6} we show the time dependence of the decay length for response and correlation in the two directions and for different shear rates. For the highest shear rate $\dot{\gamma}=10^{-4}$, $l_{R_1}$ and $l_{C_1}$ grow with time approximately in the same way, $l_x(\Delta t)\sim t^{\kappa}$ with an exponent $\kappa\approx0.5$, suggesting diffusive spreading for both the response and the correlation. Conversely, for the lower rate $\dot{\gamma}=10^{-5}$, whereas for the response the growth is still sublinear,  it's faster for the correlation, with the effect being stronger along the parallel direction, where an almost linear behavior is observed.  

The rather good agreement between response and correlation for  $\dot{\gamma}=10^{-4}$ would suggest the existence of a fluctuation-dissipation relation (FDR) $R_1(r,\Delta t) = \beta C_1(r,\Delta t)$ with $\beta$ carrying information about an effective temperature, $\beta\sim1/T_{\rm eff}$. If we restrict ourselves to short time lags $\Delta t \leqslant 20$, we estimate $\beta \sim 2.8$ for the parallel direction and $\beta \lesssim 2.0$ for the perpendicular one. The scenario is different for the lower rate $\dot{\gamma}=10^{-5}$ where response and correlation deviate more strongly and no FDR seems to hold. 

We note here that there is no theoretical justification in our system for looking a priori for a response-correlation proportionality, as the shear transformation is a strongly nonlinear local perturbation, as is the response in the form of a plastic activity. However, at a coarse grained level, and in the spirit of elasto plastic models, the shear transformation can be considered as the elementary ``dynamical event" that governs the dynamics. As a result, it is natural to investigate the similarity between the response to a triggered shear transformation (response function) and the response to one that is taking place spontaneously (correlation function). In a system at thermal equilibrium responding to a small perturbation, these two quantities are proportional, and the system does not distinguish between the external perturbation and a spontaneous fluctuations. Observing a similar property at the level of the local strain would imply that the driven system is brought, at the level of this variable,  into a state that resembles thermodynamic equilibrium at a finite temperature. 

\section{Conclusion}\label{sec:con}
This work represent an extension of a previous study \cite{PuosiSTdyn}, where we have investigated the response of a standard 2D model of glass to artificially triggered  local shear transformation which replicate  elementary plastic events observed in amorphous solids under deformation. 
No significant plastic response is observed in as-quenched configurations even for very large strains applied to the ST. By contrast, pre-sheared configurations exhibit long ranged response behavior with quadrupolar symmetry.

We have also compared quantitatively the spatiotemporal decay of the response functions to correlations between plastic events measured in the same model system during steady flow. At the highest rate considered here, correlations and response appear to be proportional to each other, suggesting the existence of a nonequilibrium generalization of the Fluctuation-Dissipation-Theorem for plastic activity. Such generalizations
that imply the existence of an effective temperature have been previously reported for other observables in driven systems during steady state \cite{BerthierTeffJCP02,OhernPRL04,HaxtonLiuPRL07}. By contrast, our (limited) data at lower shear rates suggest that this behavior does not hold in general in the present system. 

One could think of a few reasons for this difference. First, we note that we are looking at the strongly nonlinear response of a system that is not undergoing external driving (although the initial configuration has been prepared by an external drive) whereas the correlation clearly refers to a driven system in steady state. However, we find (not shown) that keeping the external drive while triggering the zone does not affect the system response. This is not surprising since we are dealing with relatively short time lags. Moreover, any effect due to stopping the driving  would be present, and probably stronger, also for the highest rate, but this seems not to be the case.

An additional possible source of discrepancy could lie by the fact that the triggered STs are instantaneous, while the spontaneous plastic events have a finite duration. From previous simulation results \cite{NicolasEPJE14}, the typical timescale of a plastic events is of the order of a few time units, which is not very well separated from the time window of the response. We expect that including a finite duration in the triggering protocol would have the effect of slowing down the response propagation, thus increasing the discrepancies with the correlation. 

Finally, the last possibility that comes to our mind is that the system around a soft spot, at which the spontaneous shear transformation is taking place, is somehow organized in a rather different manner than around the random places we are choosing to trigger the artificial transformations. This idea is consistent with the fact that FDR-like behavior is emerging at the highest driving rate, where correlations between plastic events and soft spots are reduced \cite{SchoenholzPRX14} and would deserve further attention in future work. 

\begin{acknowledgments}
The simulations were carried out using LAMMPS molecular dynamics software \cite{lammps} (\url{ http://lammps.sandia.gov}). JLB and FP are supported by Institut Universitaire de France and by grant ERC-2011-ADG20110209.  All the computations presented in this paper were performed using the Froggy platform of the CIMENT infrastructure (\url{https://ciment.ujf-grenoble.fr}), which is supported by the Rh\^one-Alpes region (GRANT CPER07\_13 CIRA) and the Equip@Meso project (reference ANR-10-EQPX-29-01) of the programme Investissements d'Avenir supervised by the Agence Nationale pour la Recherche.

\end{acknowledgments}
\bibliography{biblio_PLSTC}

\end{document}